\documentstyle[12pt]{article}

\renewcommand{\b}[2]{ b_{ {#2} }({#1})}

\newcommand{\nn}{\nonumber}

\newcommand{\bc}{\begin{center}}
\newcommand{\ec}{\end{center}}
\newcommand{\lb}{\linebreak}
\newcommand{\bq}{\begin{equation}}
\newcommand{\eq}{\end{equation}}
\newcommand{\bqa}{\begin{eqnarray}}
\newcommand{\eqa}{\end{eqnarray}}
\newcommand{\ben}{\begin{enumerate}}
\newcommand{\een}{\end{enumerate}}
\newcommand{\eqn}[1]{Eq.(\ref{#1})}
\def\q{\partial}
\def\A{\cal{A}}
\def\P{\cal{P}}
\def\l{\lambda}
\def\b{\beta}
\def\Ao#1{\A(p_1,p_2,\ldots,p_{#1})}
\def\Ad#1{\A_2(q;p_1,p_2,\ldots,p_{#1})}
\def\Aa#1#2{\A(\{p^{(#2)}_i\},#1)}
\def\Po#1#2{(\sum_{i=1}^{#1}p^{(#2)}_i)^2-m^2}
\def\Di#1#2{d^D{#1}\delta({#1}^2-{#2}^2)\theta({#1}^0)}
\def\di#1#2{d^D{#1}\delta({#1}^2-{#2}^2)\theta({#1}^0)e^{-\b\cdot {#1}}}
\def\ti#1{\tilde{#1}}
\def\d#1#2{{d^{#2}\over d{#1}^{#2}}}
\def\vphi{\varphi}

\begin{document}
\pagestyle{empty}

\begin{flushright}CERN-TH.6971/93\\
hep-ph@xxxx.lanl.gov/9309205
\end{flushright}
\vspace*{1cm}
\begin{center}\begin{Large}
{\bf
Partial-wave amplitudes \\ and multiparticle production
}\end{Large}

\vspace*{2cm}

 \vspace{\baselineskip}
{\large Costas~G.~Papadopoulos}\\
\vspace{\baselineskip}
TH Division, CERN, Geneva, Switzerland\\
\vspace*{3cm}
Abstract\\[24pt] \ec
Recursion relations for integrals of amplitudes over the
phase space, i.e. for partial wave amplitudes,
are introduced. In their simplest form
these integrals are
proportional to the s-wave amplitudes and represent
rigorous lower bounds on the total cross sections.
The connection with
classical field equations in $D$ dimensions is established.
Previous results on multiparticle amplitudes
are easily reproduced.

\vfill
\begin{flushleft} CERN-TH.6971/93\\August 1993\\
\end{flushleft}

\newpage
\pagestyle{plain}
\setcounter{page}{1}

\par
The high-multiplicity limit of processes involving scalar particles
has been studied in several contexts
\cite{gol:ampl,corn:ampl,akp:ampl,akp:cros,vol:ampl,vol:cros,rub:ampl}.
Ordinary perturbation theory
for the amplitude $\A(1\to n)$ predicts a factorial growth, $n!$,
which is inconsistent with the unitarity limits. Higher-order corrections
at threshold \cite{vol:1loo,smi:1loo,akp:loop}
exhibit an even faster growth with $n$, and the whole
perturbative expansion breaks down for sufficiently high multiplicities.
\par
On the other hand, the {\it nullification phenomenon},
i.e. the nullification
of \lb $\A(2\to n)$ amplitudes at threshold in certain theories
\cite{vol:1loo,akp:beyo,akp:nuli,bro:nuli},
including the Standard Model with specific relations
among the different couplings \cite{vol:thre,akp:unma},
might be the signal that, under certain conditions,
the perturbative expansion is still
consistent with the unitarity limits: delicate cancellations
among different graphs contributing to the amplitude
might soften the $n!$ growth below the limiting case
of the $\sqrt{n!}$ one,
so that the cross sections still satisfy unitarity bounds.
The case of sinh-Gordon interactions in two dimensions is
a quite interesting example of such cancellations, where the
nullification survives at any energy and/or at
any order of perturbation theory \cite{zam:sing}.
\vspace{\baselineskip}
\par In this paper we are introducing some new recursion relations
involving the integrals of the amplitudes over the phase space,
\bq
a(n,w)=\int \prod_{i=1}^{n}\Di{p_i}{m}\Ao{n}\delta^D(\sum p_i-P)
\label{def:a}
\eq
where $w^2=P^2$. These integrals are proportional to the
s-wave amplitudes.
Furthermore, using the well-known Schwarz inequality, we obtain
\bq
\sigma(n,w)\sim\int \prod_{i=1}^{n}\Di{p_i}{m}
|\Ao{n}|^2\ge {|a(n,w)|^2\over V_n(w)}
\label{ineq}
\eq
where $\sigma(n,w)$ is the total cross section and
$V_n(w)$ is the phase-space volume.
\par On the other hand,
recursion relations for tree-order amplitudes are, at least
formally, easily
understood. For instance the amplitude $\A(1\to n)$ can be written as
\bqa
\Ao{n}&=&\sum_{\P} \sum_{p=2}^{n} {-i\l_{p+1}\over p!}
\sum_{n_1,n_2,\ldots,n_p}
{i\Aa{n_1}{1}\over n_1!(\Po{n_1}{1})}
\nn\\
&&{i\Aa{n_2}{2}\over n_2!(\Po{n_2}{2})}\cdots
{i\Aa{n_p}{p}\over n_p!(\Po{n_p}{p})}
\label{rec:g}
\eqa
where
\bq
\Aa{n}{i}=\A(p_1^{(i)},p_2^{(i)},\ldots,p_n^{(i)})
\eq
and $\P\equiv
\{p^{(i)}_j,\;\;i=1,\ldots,p,\;\;j=1,\ldots,n_i,\;\;\sum n_i=n\}
$ represents a given permutation of the final state momenta.
The sum over $n_1,\ldots,n_p$ has to be understood under
the constraint $n_1+n_2+\cdots+n_p=n$.
The coupling constants $\l_p$ correspond to an expansion
of the interaction potential
\bq
V(\phi)=\sum_{p=3}^{\infty} {\l_p\over p!} \phi^p \;\;\;.
\eq
When all the momenta are equal and on-shell,
the kinematical configuration corresponds to the threshold case, and
the recursion relation of \eqn{rec:g}
has been studied extensively using the
well-known technique of generating functions \cite{akp:ampl}.
\par In order to combine the
recursion relation of \eqn{rec:g} with the integration
over the phase space in \eqn{def:a}, we introduce
the Laplace transform of the integrated amplitude $a(n,w)$ as
follows:
\bq
\ti{a}(n,\b)=\int d^DP e^{-\b\cdot P} a(n,w) \;\;;
\eq
which taking into account the overall $\delta$-function,
this can be
written simply as
\bq
\ti{a}(n,\b)=\int \prod_{i=1}^{n} \di{p_i}{m} \Ao{n}\;\;\;.
\eq
Defining
\bq
\ti{a}(n,\b)=-i n! (\q^2-m^2)\ti{b}(n,\b)
\;,\;\;\; \q_\mu\equiv {\q\over\q\b^\mu}
\eq
and using \eqn{rec:g}, we have
\bq
(\q^2-m^2)\ti{b}(n,\b)
= \sum_{p=2}^{n} {\l_{p+1}\over p!}
\sum_{n_1,n_2,\ldots,n_p}
\ti{b}(n_1,\b)
\ti{b}(n_2,\b)\ldots
\ti{b}(n_p,\b)\;\;\;.
\eq
We can now proceed by using the method of generating
functions. Defining that
\bq
\phi(\b;\tau)=\sum_{n=1}^{\infty} e^{-n\tau}\ti{b}(n,\b)
\label{genf:def}\eq
we find that the generating function satisfies the classical
field equations given by
\bq
(\q^2-m^2)\phi=V'(\phi) \;\;\;. \label{dif:dd}
\eq
The above equation can be seen as a generalization
of the well-known result that the generating function
of tree amplitudes satisfies the classical field
equations \cite{bro:clas}: integrals of amplitudes over the
phase space can be calculated as well using
the classical solutions in $D$ dimensions.
On the other hand, since we seek only solutions
that depend on the measure of the vector $\b^\mu$, \eqn{dif:dd}
becomes an ordinary (non-linear) differential equation
\bq
\Bigl(\d{\b}{2}+{D-1\over \b}\d{\b}{}-m^2\Bigr)\phi=V'(\phi)
\label{dif:1d}
\eq
which under certain circumstances can be solved exactly.
\par However hand we have to be careful
when we consider \eqn{dif:1d}
in connection with \eqn{genf:def}. Since we are dealing with a
system of differential equations the answer might not be unique.
This fact is reflected in the dependence of the solutions of
\eqn{dif:1d} on $\tau$, which will indroduce an `infinite'
arbitrariness. This can be illustrated in the following example.
Let us choose $b_n(x)=ne^{-nx}$, which is the Laplace transform
of a $\delta$-function, $\delta(t-n)$. They satisfy
the following system of ordinary differential equations
\bq
\Bigl(\d{x}{2}-1\Bigr) b_n(x)=6\sum_{n_1,n_2}b_{n_1}(x)b_{n_2}(x)
\label{dif:exa}
\eq
which can be seen as a one-dimensional $\phi^3$ theory.
Defining $\phi=\sum_{n=1}^{\infty} e^{-n\tau}b_n(x)$
\eqn{dif:exa} becomes
\bq
\Bigl(\d{x}{2}-1\Bigr)\phi=6\phi^2 \eq
with the solution
\bq \phi(x;\tau)={1\over 4}\sinh^{-2}\Bigl({x+x_0(\tau)\over 2}
\Bigr)\;\;\;.
\eq
It is now obvious that the existence of an arbitrary function
of $\tau$, $x_0(\tau)$,
does not allow to calculate the `amplitudes' $b_n$ unless
some other conditions are specified. For instance, if we require
that the inverse Laplace transform of $b_n(x)$ vanish for
$t < n$, we single out the initial solution. Indeed
this requirement applies also in the case of $D > 1$, since all
physical amplitudes are non-zero only for $w\ge nm$.
\par Having the solution of \eqn{dif:1d}, we can find the integrated
amplitude as follows \cite{ron:phsp},
\bq
a(n,w)=w^{1-D/2} {\pi^{1/2}\over 2^{D/2-1}\Omega_D\Gamma({D-1\over 2})}
{1\over i\pi}\int_{c-i\infty}^{c+i\infty}d\b\;
I_\nu(\b w)\;\b^{D/2}\;\ti{a}(n,\b)
\label{inv}\eq
where $\nu=D/2-1$, $I_\nu$ is the modified
Bessel function and
\bq
 \Omega_D=\left\{ \begin{array}{ll}1 & \mbox{if $D=2$}
\\ 2\pi^{D-1\over 2}/
\Gamma({D-1\over 2}) & \mbox{if $D\ge 3$} \;\;\;. \end{array}
\right.
\eq
\vspace{\baselineskip}
\par Several results can be reproduced using the above formalism.
First of all expanding around $\b=\infty$, which in $w$-space
corresponds to an expansion around threshold, $w=nm$, we get
($D=4$)
\bq
b(n,\b)=e^{-n\b m} \b^{-3(n-1)/2}\sum_{k=0}^{\infty}
b_k(n) \b^{-k} \;\;\;.
\eq
The first coefficient, $b_0(n)$, satisfies the
equation
\bq
(n^2-1)b_0(n)
= \sum_{p=2}^{n} {\l_{p+1}\over p!}
\sum_{n_1,n_2,\ldots,n_p} b_0(n_1)b_0(n_2)\ldots b_0(n_p)
\label{rec:cl}
\eq
with the well-known result \cite{akp:ampl,akp:beyo}.
Notice that the leading
term in the expansion around $w=nm$, is proportional to
$n!(w-nm)^{3n-5\over 2}/\Gamma({3\over 2}(n-1))$
which, taking $(w-nm)$ to be constant, is not unitarity-violating
as we expected, since it corresponds to the non-relativistic
approximation. In contrast, the saddle-point approximation for
the integral in \eqn{inv} should be used when $w\to \infty$, but
$f=w/nm$ is constant. Assuming that in this limit,
$b(n,\b)\sim b_0(n)\exp(-n\vphi(\b))$, the saddle-point equation
is given by
\bq
\d{\b}{}\vphi(\b)=f \;\;\;. \label{sad:1}
\eq
This leads to the following
recursion relation for the leading coefficient
\bq
f^2(n^2-1)b_0(n)=
\sum_{p=2}^{n} {\l_{p+1}\over p!}
\sum_{n_1,n_2,\ldots,n_p} b_0(n_1)b_0(n_2)\ldots b_0(n_p)
\label{rec:lol}
\eq
where we have replaced $n^2$ by $(n^2-1)$ in order to make the
recursion relation consistent for all $n\ge 1$. It is easy to
see that \eqn{rec:lol}
is nothing but the recursion relation for the
lower bound on the amplitude given in references
\cite{akp:ampl,akp:cros} and
that $\vphi(\b)\sim \log(K_1(\b m)/\b)$, which
is given by the initial condition that
\bq b(1,\b)={2\pi m\;K_1(\b m)\over \b}\;\;\;. \eq
\par Another advantage of our method is that one can study massless
theories as well, for which the threshold amplitude is
singular. This is possible because,
in several cases, the singularity of the
propagator is cancelled by the phase-space integration.
For instance, for the
$\phi^4$ in four space-time dimensions, it is easy to show
that the result is indeed finite for the integrated amplitude
and that \eqn{rec:g} is suitable for the study of multiparticle
massless amplitudes.
\vspace{\baselineskip}
\par The amplitude $\A(2\to n)$ can be studied in exactly
the same way.
The resulting equation is given by
\bq
(\q^2+2q\cdot\q)\ti{c}(n,\b,y)
= \sum_{p=1}^{n} {\l_{p+2}\over p!}
\sum_{n_1,n_2,\ldots,n_p}
\ti{c}(n_1,\b,y)
\ti{b}(n_2,\b)\ldots
\ti{b}(n_p,\b)
\label{2ton:dd}
\eq
where $y=\b\cdot q/\b$ and
\bq
-in!(\q^2+2q\cdot\q)\ti{c}(n,\b,y)=\int d^DP e^{-\b\cdot P}
a_2(n,w,P\cdot q) \label{def:lapt}
\eq
with
\bq
a_2(n,w,P\cdot q)=
\int \prod_{i=1}^{n}\Di{p_i}{m}\Ad{n}\delta^D(\sum p_i-P)
\eq
where $\Ad{n}$ is the amplitude for the process
\[\phi(q')+\phi(q)\to \phi(p_1)+\cdots+\phi(p_n)\;\;\;.\]
Of course
for the physical amplitude we have to take the limit
$q'^2\to m^2$, or $2 P\cdot q\to w^2$, so that there is a
one-to-one correspondence between the physical amplitudes
and the poles of the function $c(n,w,P\cdot q)$ in $(w^2-2P\cdot q)$.
\vspace{\baselineskip}
\par In order to study \eqn{2ton:dd},
we expand around threshold, which in $\b$-space
corresponds to an expansion around $\b=\infty$. The leading
coefficient satisfies the following equation
\bq
n(n-2E)c_0(n,E)=
\sum_{p=1}^{n} {\l_{p+2}\over p!}
\sum_{n_1,n_2,\ldots,n_p}
c_0(n_1,E)b_0(n_2)\ldots b_0(n_p)
\label{nul:cl}
\eq
where $b_0$ satisfies \eqn{rec:cl}. Equation (\ref{nul:cl}) exhibits
the nullification properties, i.e. the existence
of a finite number of poles in $(n-2E)$ for certain theories
\cite{akp:beyo,akp:nuli}. The
next-to-leading coefficient, $c_1(n,E)$, satisfies the equation
$(D=2)$:
\bqa
&& n(n-2E)c_1(n,E)+(n^2-n-nE)c_0(n,E)+2(1-E^2)\d{E}{}c_0(n,E)\nn\\
&& = \sum_{p=1}^{n} {\l_{p+2}\over p!}
\sum_{n_1,n_2,\ldots,n_p}
( c_1(n_1,E)b_0(n_2)+p c_0(n_1,E)b_1(n_2) )\ldots b_0(n_p)
\label{rec:21}
\eqa
where $b_1(n)$ satisfies the equation
\bq
(n^2-1)b_1(n)+n(n-1)b_0(n)=
\sum_{p=2}^{n} {\l_{p+1}\over p!}
\sum_{n_1,n_2,\ldots,n_p}
b_1(n_1)b_0(n_2)\ldots b_0(n_p) \;\;\;.
\eq
\par As a check of these equations
we consider the case of the sinh-Gordon
potential in two space-time dimensions, which is given by
(including the mass term)
\bq
U(\phi)={m^2\over \l^2}(\cosh(\l\phi)-1) \;\;\;.
\eq
The result of the recursion relation \eqn{rec:21}, taking for simplicity
$m=1$, $\l=1$ so that $\l_{p+1}=0$ for $p$ even and
$\l_{p+1}=1$ for $p$ odd,
is the absence of any pole other than $E=1$
for $c_1(n,E)$, as expected, since we know that the nullification
for this model is exact for any energy and not only at threshold
\cite{zam:sing}.
This is a nice test for the validity of our formalism.

\par Let us now study the high-$n$ behaviour of the function
$\ti{c}(n,\b,y)$ in the limit $n\to \infty$, $w\to \infty$,
$f$=$w/nm$=constant. This can be done by studying the inverse
Laplace transform in \eqn{def:lapt}. The saddle-point equation
in $\b$-space can be written as:
\bq
 \q_\mu F(\b,y)=P_\mu \eq
where $F\sim -\log(\ti{c})$. The solution is given
by $F=n\vphi(\b)$, where $\varphi$ satisfies \eqn{sad:1}.
Writing $\ti{c}=c_0(n,E)\exp(-n\vphi(\b))$, and
keeping the leading terms in $n$, we find that
\bq
(f^2n^2-2fnE)c_0(n,E)=
\sum_{p=1}^{n} {\l_{p+2}\over p!}
\sum_{n_1,n_2,\ldots,n_p}
c_0(n_1,E)b_0(n_2)\ldots b_0(n_p)
\eq
where we consider $E$ to be of order $n$. It is interesting
to notice that if $b_0$ satisfies \eqn{rec:lol}
then $c_0(n,E)$ still has a finite number of poles
in $(nf-2E)$ for
the theories with the nullification property.
The only difference from the
threshold case is a trivial redefinition of the
couplings, $\l\to \l/f^2$, and the fact
that the poles have been shifted
from $E=n/2$ to $E=fn/2$, as they should.
It is also interesting to notice that
this result does not depend on the dimensionality of the space-time,
which enters only in subleading terms, and that it is indeed
true for the case of the sinh-Gordon potential in two space-time
dimensions.
Although the question of nullification beyond threshold
is of capital importance, a complete study of this
problem in the context of 4-dimensional theories is beyond the scope
of this paper and will be given elsewhere \cite{akp:prep}.
\vspace{\baselineskip}
\par We conclude by emphasizing that the formalism we
propose in this paper leads to recursion relations for
integrated (partial-wave) amplitudes,
which can be used to reconstruct the
total cross section. The simple integral of the amplitude
over the phase space, which is proportional to the s-wave amplitude,
is a rigorous lower bound on the total cross section.
This formalism is quite general
and has the very interesting property
that the generating function of the Laplace-transformed
integrated amplitudes, at tree order, satisfies
the classical equation of motion in $D$ dimensions.
All previously known results can easily be reproduced,
using appropriate approximation methods.
Furthermore, this formalism enables us to directly relate
tree-order unitarity, in the limit of high multiplicities,
with the classical field equations, and offers the unique
possibility to study multiparticle production above
threshold.
\vspace*{2cm}
\section*{Acknowledgements}
I wish to thank E.~N.~Argyres and R.~Kleiss for helpful discussions.
I am also grateful to Mrs Susy Vascotto for her assistance in
preparing the manuscript.

\end{document}